
\titlepage
\baselineskip=24pt
\hsize=6.5in
\line{DESY 93-131\hfill ISSN 0418-9833}
\line{MAD/TH/93-07\hfill}
\line{September 1993\hfill}
\vskip1cm
\title{NEW PERTURBATIVE UPPER BOUND ON $M_H$  FROM FERMIONIC
HIGGS DECAYS AT TWO LOOPS}
\author{Loyal Durand,$\!^*$ Bernd A. Kniehl,$\!^\dagger$ and
Kurt Riesselmann$^*$}
\vskip1cm
\line{\hskip3cm$^*$ Department of Physics, University of Wisconsin,
\hfill}
\line{\hskip3cm\phantom{$^*$} 1150 University Avenue, Madison, WI 53706, USA
\hfill}
\line{\hskip3cm$^\dagger$ II. Institut f\"ur Theoretische Physik,
Universit\"at Hamburg,\hfill}
\line{\hskip3cm\phantom{$^\dagger$} Luruper Chaussee 149, 22761 Hamburg,
Germany\hfill}
\vfil
\abstract
We present the dominant two-loop ${\rm O}\left(G_F^2M_H^4\right)$
electroweak corrections to the fermi\-onic decay widths of a
high-mass Higgs boson in the Standard Model.
The corrections are negative and quite significant, and are
larger in magnitude than
the one-loop electroweak corrections
for $M_H\gsim 400\ {\rm GeV}$.
This indicates the onset of a breakdown of perturbation theory
in the Higgs sector of the Standard Model
at this surprisingly low value of the Higgs-boson mass.
\endpage
\def\CPC #1 #2 #3 {Comp.\ Phys.\ Commun.\ {\bf #1}\ (#2) #3}
\def\FP #1 #2 #3 {Fortschr.\ Phys.\ {\bf #1} (#2) #3}
\def\IJMP #1 #2 #3 {Int.\ J.\ Mod.\ Phys.\ {\bf #1}\ (#2) #3}
\def\MPL #1 #2 #3 {Mod.\ Phys.\ Lett.\ {\bf #1}\ (#2) #3}
\def\NP #1 #2 #3 {Nucl.\ Phys.\ {\bf #1},\ #2\  (#3)}
\def\NC #1 #2 #3 {Nuovo Cim.\ {\bf #1} (#2) #3}
\def\PL #1 #2 #3 {Phys.\ Lett.\ B\ {\bf #1},\ #2 (#3)}
\def\PP #1 #2 #3 {Phys.\ Rep.\ {\bf #1}\ (#2) #3}
\def\PR #1 #2 #3 {Phys.\ Rev.\ D\ {\bf #1},\ #2 (#3)}
\def\PRL #1 #2 #3 {Phys.\ Rev.\ Lett.\ {\bf #1},\ #2 (#3)}
\def\RMP #1 #2 #3 {Rev.\ Mod.\ Phys.\ {\bf #1}\ (#2) #3}
\def\ZP #1 #2 #3 {Z.\ Phys.\ C\ {\bf #1},\ #2 (#3)}
%
\REF\sch{J. Schwindling, in {\it Proceedings of the International
Europhysics Conference on High Energy Physics},
Marseille, France, July 22--28, 1993, ed.\ by J. Carr and M. Perrottet
(Editions Fronti\`eres, Gif-sur-Yvette, 1993) to appear.}
\REF\dic{D.A. Dicus and V.S. Mathur, \PR 7 3111 1973  .}
\REF\lee{B.W. Lee, C. Quigg, and H.B. Thacker, \PRL 38 883 1977 ;
\PR 16 1519 1977 .}
\REF\vel{M. Veltman, Acta Phys.\ Pol.\ {\bf B8} (1977) 475.}
\REF\mve{M. Veltman, Phys.\ Lett. {\bf 70B}, 253 (1977).}
\REF\unit{L. Durand, P.N. Maher, and K. Riesselmann, \PR 48 1084 1993 .}
\REF\has{A. Hasenfratz and T. Neuhaus, \NP B297 205 1988 ;
W. Langguth and I. Montvay, \ZP 36 725 1987 ;
A. Hasenfratz, T. Neuhaus, K. Jansen, H. Yoneyama, and C.B. Lang,
\PL 199 531 1987 ;
P. Hasenfratz and J. Nager, \ZP 37 477 1988 ;
M. L\"uscher and P. Weisz, \PL 212 472 1988 ;
U.M. Heller, H. Neuberger, and P. Vranas, \NP B399 271 1993 .}
\REF\kni{B.A. Kniehl, Report No.\ DESY 93-064 (August 1993),
Phys.\ Rep.\ (to appear).}
\REF\lem{M. Lemoine and M. Veltman, \NP B164 445 1980 .}
\REF\hal{J. van der Bij and M. Veltman, \NP B231 205 1984 ;
J.J. van der Bij, \NP B248 141 1984 ;
F. Halzen and B.A. Kniehl, \NP B353 567 1991 ;
F. Halzen, B.A. Kniehl, and M.L. Stong, \ZP 58 119 1993 .}
\REF\mar{W.J. Marciano and S.S.D. Willenbrock, \PR 37 2509 1988 .}
\REF\hvv{J. Fleischer and F. Jegerlehner, \PR 23 2001 1981 ;
B.A. Kniehl, \NP B352 1 1991 ; {\it ibid.} B357 439 (1991) .}
\REF\hff{D.Yu.~Bardin, B.M. Vilenski\u\i{}, P.Kh.~Khristov,
Yad.\ Fiz.\ {\bf53}, 240 (1991)
[Sov.\ J.\ Nucl.\ Phys.\ {\bf53}, 152 (1991)];
B.A. Kniehl, \NP B376 3 1992 ;
A. Dabelstein and W. Hollik, \ZP 53 507 1992 .}
\REF\pro{J. Fleischer and F. Jegerlehner, \NP B216 469 1983 ;
B.A. Kniehl, \ZP 55 605 1992 ;
A. Denner, J. K\"ublbeck, R. Mertig, and M. B\"ohm, \ZP 56 261 1992 .}
\REF\det{Full details will be presented in a separate communication.}
\REF\mah{P.N. Maher, L. Durand, and K. Riesselmann, \PR 48 1061 1993 .}
\REF\cor{J.M. Cornwall, D.N. Levin, and G. Tiktopoulos, \PR 10 1145 1974 ;
 {\bf D11}, 972(E) (1975) ;
C.E. Vayonakis, Lett.\ Nuovo Cim.\ {\bf 17}, 383 (1976);
M.S. Chanowitz and M.K. Gaillard, \NP B261 379 1985 ;
G.J. Gounaris, R. K\"ogerler, and H. Neufeld, \PR 34 3257 1986 ;
J. Bagger and C. Schmidt, \PR 41 264 1990 ;
H. Veltman, \PR 41 2294 1990 ; H.-J. He, Y.-P. Kuang, and
X. Li, \PRL 69 2619 1992 .}
\REF\bee{W. Beenakker and W. Hollik, \ZP 40 141 1988 .}
\REF\uni{W.W. Repko and C.J. Suchyta, \PRL 62 859 1989 ;
D.A. Dicus and W.W. Repko, \PL B228 503 1989 ; \PR 42 3660 1990 ;
G. Valencia and S. Willenbrock, \PR 42 853 1990 ;
H. Veltman and M. Veltman, Acta Phys.\ Pol.\ {\bf B22}, 669 (1991);
K. Hikasa and K. Igi, \PL 261 285 1991.}
%

One of the great open questions of elementary particle physics today
is whether nature makes use of the Higgs mechanism of
spontaneous symmetry breaking to generate the observed particle masses.
The Higgs boson, $H$, is the missing link sought to verify
this concept in the Standard Model (SM).
Many of the properties of the Higgs boson are fixed, \eg,
its couplings to the gauge bosons,
$g_{VVH}=2^{5/4}G_F^{1/2}M_V^2\ (V=W,Z)$, and fermions,
$g_{f\bar fH}=2^{1/4}G_F^{1/2}m_f$,
and the vacuum expectation value,
$v=2^{-1/4}G_F^{-1/2}\approx246\ {\rm GeV}$.
However, its mass, $M_H$, and its self-couplings, which depend on $M_H$,
are essentially unspecified.

The failure of experiments at LEP~1 and SLC to detect the decay
$Z\rightarrow f\bar f H$ has ruled out the mass
range $M_H\leq $ 63.5~GeV at the 95\%
confidence level [\sch].
At the other extreme, unitarity arguments in intermediate-boson scattering
at high energies [\dic,\lee] and considerations concerning the range of
validity of perturbation theory [\vel,\mve] establish an upper bound on
$M_H$ at $(8\pi\sqrt2/3G_F)^{1/2}\approx1{\rm TeV}$ in a weakly interacting SM.
This bound is lowered significantly when the approach of [\dic,\lee] is
extended to higher orders (see [\unit] and references therein).
However, the bounds obtained depend on the energy scale up to which the SM is
supposed to be valid. Independently,
lattice computations [\has] suggest an upper bound on $M_H$ of $\sim$
630~GeV.

The main result of the present letter will be that, in the
case of the fermionic decays of the Higgs boson, the two-loop electroweak
radiative corrections to the decay rates already exceed the one-loop
corrections in magnitude at $M_H\approx400\ {\rm GeV}$.
The perturbation expansion therefore fails to converge satisfactorily.
This result establishes
a new perturbative upper bound on $M_H$:
if the SM is to be weakly interacting so that perturbation theory is
meaningful, $M_H$ must be less than 400 GeV.
Note that this bound is independent of assumptions concerning the
energy scale up to which the SM is valid. A more
massive Higgs boson will necessarily be strongly interacting.

Radiative corrections to the production and decay rates of the Higgs boson
have received much attention in the literature; for a recent review,
see [\kni].
It is well known that, in $Z$-boson physics, the quantum effects induced by
virtual Higgs bosons are screened [\vel,\lem]. They are logarithmic at one
loop and quadratic, but with minute coefficients, at two loops [\hal].
By contrast, the one-loop electroweak corrections to the partial decay widths
[\vel,\mar--\hff] and production cross sections [\pro] of the Higgs boson
are dominated for $M_H\gg M_W$ by terms of ${\rm O}\left(G_FM_H^2\right)$.
In all one-loop calculations considered, these terms give rise to
enhancements of the rates which are moderate for $M_H\lsim1{\rm TeV}$.
However, it is premature to conclude from this observation that the
two-loop electroweak corrections are also perturbatively small for
$M_H$ large.
It is both of theoretical and phenomenological interest to check
this point by explicit calculation.
In this letter, we present the results of a first step towards this
goal, the calculation of
the leading two-loop electroweak corrections to the fermionic decay
rates of a heavy Higgs boson with $M_H\gg M_W$.
The corrections are ${\rm O}\left(G_F^2M_H^4\right)$, and
are independent of the fermion flavour.

We briefly outline the basics of our calculation [\det],
which largely proceeds along the lines of [\mah].
We work in the electroweak on-mass-shell renormalization
scheme and calculate the
${\rm O}\left(G_FM_H^2\right)$ and ${\rm O}\left(G_F^2M_H^4\right)$
corrections to $\Gamma\left(H\rightarrow f\bar f\,\right)$.
In the limit of $M_H\gg M_W$, power counting and
inspection of coupling constants
(in the 't~Hooft-Feynman gauge) reveal that we may concentrate on those
one- and two-loop diagrams which involve only the physical Higgs boson
and the longitudinal polarization states of the intermediate bosons.
The fermion mass and wave-function renormalizations do not receive
contributions in the orders considered.
Furthermore, we may safely neglect the gauge couplings and
intermediate-boson masses.
Consequently, we may apply the Goldstone-boson equivalence theorem [\cor].
All necessary information may be obtained from the Lagrangian of
the symmetry-breaking sector, which characterizes the kinematics and
interactions of the Higgs boson, $H$, and the Goldstone bosons,
$w^\pm$ and $z$.
The latter remain massless and satisfy a residual SO(3) symmetry.
The boson masses and wave functions are renormalized according to the
usual on-mass-shell procedure.
Futhermore, we fix the physical vacuum expectation value and quartic coupling
by $v=2^{-1/4}G_F^{-1/2}$ and $\lambda=G_FM_H^2/\sqrt{2}$, respectively,
where $G_F$ is the Fermi constant.
Straightforward algebra shows that the Higgs-fermion Yukawa coupling
receives a multiplicative correction of the form $(Z_H/Z_w)^{1/2}$,
where $Z_H$ and $Z_w$ are the wave-function renormalizations of
$H$ and $(w^\pm,z)$.
Thus, the leading high-$M_H$ corrections to the fermionic Higgs-boson
decay rates can be included by multiplying the tree-level result
for $\Gamma\left(H\rightarrow f\bar f\,\right)$ by the overall factor
$Z_H/Z_w$,
independently of the fermions involved.

The wave-function renormalizations $Z_H$ and $Z_w$ are defined as [\mah]
$$
{1\over Z_w}=1-\left.{d\over dp^2}\Pi_w^0(p^2)\right|_{p^2=0}\,,
\qquad
{1\over Z_H}=1-\left.{d\over dp^2}{\ rm Re}\Pi_H^0(p^2)\right|_{p^2=M_H^2}\,,
\eqno\eq$$
where $\Pi_w^0(p^2)$ and $\Pi_H^0(p^2)$ are the self-energy functions for
the bare fields.
Explicit expressions may be found in Eqs.~(11), (12) of [\mah].
Using dimensional regularization, we find that $Z^{-1}$ can be written
in factored form,
$$
	{1\over Z_\sigma}=
	\left( 1 + a_\sigma\hat\lambda + b_\sigma\hat\lambda^2  \right)
	\left( 1 + {3\over\epsilon}\hat\lambda^2\right)
	+ {\rm O}\left(\hat\lambda^3\right)\,, \qquad \sigma = w, H\,,
\eqn\eqzw$$
where $\hat\lambda=(\lambda/16\pi^2)$, $a_w=1$,
$a_H=2\pi\sqrt3-12\approx -1.12$,
$b_w\approx-24.76$, and $b_H\approx265.88$ [\det].
Hence,
$${Z_H\over Z_w}={1+a_w\hat\lambda+b_w\hat\lambda^2\over
1+a_H\hat\lambda+b_H\hat\lambda^2}\,.
\eqn\eqres$$
Note that while $Z_H$ and $Z_w$ are separately
plagued by ultraviolet singularities at ${\rm O}
(\hat\lambda^2)$ for $\epsilon\rightarrow 0$,
the singular factors cancel in the ratio as they must, since
$\Gamma\left(H\rightarrow f\bar f\,\right)$ is a physical quantity.
Equation~{\eqres}, which naturally emerges from our formalism,
automatically resums one-particle-reducible Higgs-boson self-energy
diagrams in a way that conforms with the standard procedure
in $Z$-boson physics; see, \eg, [\bee].
Strictly speaking, we have no control of terms beyond ${\rm O}(\hat\lambda^2)$.
Expanding Eq.~{\eqres} and discarding terms beyond ${\rm O}
(\hat\lambda^2)={\rm O}(G_F^2M_H^4)$,
we obtain the alternative representation
$$
{Z_H\over Z_w} = 1+(a_w-a_H)\hat\lambda+
\left(b_w - b_H - a_wa_H + a_H^2\right)\hat\lambda^2\,,
\eqn\eqexp$$
which extends the well-known one-loop result [\mve,\mar],
$$
{Z_H\over
Z_w}=1+{G_FM_H^2\over8\pi^2\sqrt2}\left({13\over2}-\pi\sqrt3\,\right).
$$

We are now in a position to explore the implications
of our results.
\FIG\one{Universal electroweak correction factor for
$\Gamma\left(H\rightarrow f\bar f\,\right)$ to ${\rm O}\left(G_FM_H^2\right)$
and ${\rm O}\left(G_F^2M_H^4\right)$ in the windows (a) $100\ {\rm GeV}
\le M_H\le 550\ {\rm GeV}$
and (b) $100\ {\rm GeV}\le M_H\le 1000\ {\rm GeV}$.
In each order, the expanded result, Eq.~{\eqexp}, is compared
with the calculation where the one-particle-reducible Higgs-boson
self-energy diagrams are resummed, Eq.~{\eqres}.
The inclusion of the two-loop term quenches the correction
at $M_H=375\ {\rm GeV}$ and flips its sign at $M_H=530\ {\rm GeV}$.}
In Fig.~{\one}, we show the leading electroweak
corrections to $\Gamma\left(H\rightarrow f\bar f\,\right)$ in the one- and
two-loop approximations with and without resummation of
one-particle-reducible higher-order terms plotted as functions of $M_H$.
Let us first concentrate on the expanded expression in Eq.~{\eqexp}.
While the ${\rm O}\left(G_FM_H^2\right)$ term (upper dotted line)
gives a modest increase in the rates, by 11\% at $M_H=1{\rm TeV}$,
the situation changes drastically when the two-loop term is included.
In fact, already at $M_H=375\ {\rm GeV}$,
it fully compensates the one-loop term.
At $M_H=530\ {\rm GeV}$, the sum of one- and two-loop corrections
(lower dotted line) reaches the same magnitude as the one-loop corrections,
but with a reversed sign.
The perturbation series for $\Gamma\left(H\rightarrow f\bar f\,\right)$
clearly ceases to converge usefully, if at all,
for values of $M_H$ beyond 400~GeV.
A Higgs boson with a mass larger than 400~GeV effectively becomes a
strongly interacting particle, a very surprising result. Conversely,
$M_H$ must not exceed approximately
400~GeV if the standard electroweak perturbation theory is to be predictive.
Note that, for $M_H\gsim$ 400~GeV, one cannot use the usual unitarization
schemes invoked in studies of $W_L^\pm,Z_L,H$ scattering [\lee,\uni]
to restore the predictiveness for the Higgs-boson width, as no
unitarity violation occurs.

We may expect to improve our perturbative result in the upper $M_H$
range somewhat
by resumming the one-particle-reducible contributions to the Higgs-boson
wave-function renormalization according to Eq.~{\eqres}.
This leads to an insignificant increase of the one-loop correction
(upper solid line), while the negative effect of the two-loop correction
is appreciably lessened (lower solid line), \ie,
the ratio of two- to one-loop corrections is rendered more favorable
theoretically.
However, in the mass range below $M_H=600\ {\rm GeV}$, this effect is too
feeble
to change our conclusions concerning the breakdown of perturbation theory
and its implications for the upper bound on $M_H$.

The results above are based on the use of the Goldstone-boson
equivalence theorem and the neglect of ${\rm O}(g^2)$ terms in the gauge
couplings.  This is expected to be an excellent approximation in the case of
$M_W/M_H \propto gv/M_H \ll 1$.
To get an idea how large the unknown subleading two-loop electroweak
corrections are for $M_H\leq1$ TeV,
we have compared the one-loop ${\rm O}\left(G_FM_H^2\right)$ term with
the full one-loop electroweak corrections [\hff],
and find that the leading correction to the $H\rightarrow t\bar t$
decay width underestimates the full
correction by 34\% (21\%) at $M_H=500\ {\rm GeV}$ (1~TeV).
This suggests that the leading terms at two loops,
${\rm O}\left(G_F^2M_H^4\right)$, will also give a
reasonable estimate of the
full two-loop result. Certainly, the subleading terms will not restore
the predictiveness of the SM for $M_H$ large.
Other subleading two-loop corrections to the fermionic Higgs-boson width
depend on the fermionic channel considered.
Since the coupling of the Higgs boson to fermions is different for each
flavor, it is improbable that the non-perturbative behaviour of
the universal correction $Z_H/Z_w$ for large $M_H$
could be cancelled in all channels simultaneously.

In summary, we have calculated the leading two-loop electroweak
corrections to the fermionic decay rates of a high-mass Higgs boson
in the SM, which are of ${\rm O}\left(G_F^2M_H^4\right)$.
They are negative and already exceed in magnitude the positive one-loop
${\rm O}\left(G_FM_H^2\right)$ corrections for $M_H\approx$ 400~GeV.
We conclude that this value has to be considered as a new theoretical
upper bound on $M_H$ in a predictive electroweak perturbation theory.
We emphasize that this argument is independent of speculations regarding
the energy scale up to which the SM is valid.

One of us (BAK) would like to express his gratitude to the Physics
Department of UW-Madison for supporting his visit, during which part of
this work was carried out, and for the great hospitality extended to him.
This work was supported in part by the U.S. Department of Energy under
Contract No.\ AC02--76ER00881.
\endpage
\refout
\endpage
\figout
\bye